\documentclass[12pt]{book}
\usepackage{plenum}
\usepackage{latexsym}
\usepackage{epsfig}
\input psfig
\newcommand{\msbar}{{\rm \overline{MS\kern-0.14em}\kern0.14em}}
\newcommand{\site}[1]{\refnote{\cite{#1}}}
\begin{document}
\chapter{HADRONIC PHYSICS FROM THE LATTICE}

\author{Chris Michael}

\affiliation{Theoretical Physics Division, 
Department of Mathematical Sciences, \\
University of Liverpool, L69 3BX, UK
}

\section{INTRODUCTION}

Topics covered in this review are the lattice gauge theory approach to
the evaluation of  non-perturbative hadronic interactions from first
principles, particularly applications to glueballs, inter-quark
potentials, the running  coupling constant and hybrid mesons. Also
discussed are the limitations of  the quenched approximation.

The theory of the strong interactions is accurately provided by Quantum
Chromodynamics (QCD). The theory is defined in terms  of elementary
components: quarks and gluons. The only free parameters  are the quark
mass values (apart from an overall energy scale imposed  by the need to
regulate the theory). This formulation is essentially the unique
candidate for the theory of the strong interactions. The only feasible
way to  describe a departure from QCD would be in terms of quark and
gluon  substructure.  At least at the energy scales up to which it is
tested, there  is no evidence for such substructure.

QCD provides a big challenge to theoretical physicists. It is defined 
in terms of quarks and gluons but the physical particles are 
composites: the mesons and baryons. Any complete description must  then
yield these bound states: this requires a non-perturbative  approach.
One can see the limitations of a perturbative approach by  considering
the vacuum: this will be approximated in perturbation theory as
basically empty with  rare quark or gluon loop fluctuations. Such a
description will allow  quarks and gluons to propagate essentially
freely which is not the case experimentally. The true (non-perturbative)
vacuum can be better thought of as a  disordered medium with whirlpools
of colour on different scales. Such a  non-perturbative treatment then
has the possibility to explain  why quarks and gluons do not propagate
(i.e. quark confinement). Furthermore it will be able to explore other 
situations than experiment --- different quark masses, etc.  My notation
is that there are $N_f$  flavours of light quark  with $N$ colours.

Any mathematically precise description of QCD must introduce some
regularisation of the  ultra-violet divergences. Any such regularisation
may spoil the symmetries,  for example  4-dimensional  Lorentz invariance
is broken by dimensional regularisation. Introducing a space time
lattice likewise  breaks Lorentz invariance. Using a space time lattice
with exact gauge invariance retained  proves to be a very successful
regularisation. In the continuum  limit, as the lattice spacing is
reduced to zero, Lorentz invariance  will be found to be restored.

For a general review of lattice gauge theory, see standard
textbooks\site{texts}. Here I give a brief introduction to the salient
points.   The simplest  discretisation of space-time is that introduced 
by Wilson\site{wilson}. Space-time is replaced by a discrete grid (the
lattice) but gauge invariance is  retained exactly. Then gluonic colour
fields  which relate the colour coordinate system at different space
time points  will be represented as links on this lattice. Thus a link
of the lattice from $x$ to $x+a$ in the $\mu$-direction (with unit
vector $e_{\mu}$) is represented by a $N \times N$ colour matrix which
can be  thought of as the path-ordered exponential of the  continuum
colour fields $A_{\mu}(x)$,
 \begin{equation}
 U_{\mu}(x) \approx e^{ igaA_{\mu}(x+{1 \over 2}ae_{\mu}) }.
 \end{equation}
 From these link matrices, the simplest non-trivial gauge invariant is
the  trace of the path-ordered product of links around a unit square:
the plaquette.
 \begin{equation}
  S_{\Box} = { 1 \over N} {\rm Re Tr}(\prod U).
 \end{equation}
 Note that, with this definition, a cold lattice with $A \approx 0$ will
have all link matrices $U \approx 1$ and hence have $S_{\Box} \approx
1$ , while  a hot lattice with random $U$ will have  $S_{\Box} \approx
0$.
 This gauge invariant $S_{\Box}$, when summed over all space-time, is
the simplest candidate  for the gluonic component of the action: the
Wilson gauge action. This follows since naively (i.e. neglecting quantum
 fluctuations) as $a \to 0$
 \begin{equation}
   \beta \sum_{\rm hypercube} S_{\Box}-1 \  \to \ -{1 \over 4} \int
F_{\mu \nu} F_{\mu \nu} \, d^4 x 
 \end{equation} 
 where $\beta= 2N / g^2$.

 Using  periodic boundary conditions in space and  time, the system 
will have a finite number of degrees of freedom:  the gluon and quark
fields at the lattice sites. This finite number of degrees of freedom
implies that the theory is a quantum many-body problem rather than a
field theory.  The step which makes this many-body problem tractable is
to consider  Euclidean time.  With the Euclidean time approach, the
formulation of QCD (consider, for  example, the  functional integral
over the gauge fields) is converted into a  multiple integral which is
well defined mathematically and has  a positive definite integrand:
 \begin{equation}
    {\cal Z} = \int DU e^{\beta \sum S_{\Box}},
 \end{equation}
 where the integral is over the group manifold of $SU(N)$ of colour for
each  link matrix $U$. Because of gauge invariance not all links are
independent, but integrating over the dependent links only introduces a
finite constant which is irrelevant. For a  lattice of $L^4$ sites with
a colour gauge group of SU(3), this  would be a $8 \times  4 \times L^4$
dimensional integral (8 from  the colour group manifold, $4L^4$ from the
gauge fields on each link). For any reasonable value of $L$,  this is a
very high dimension indeed. Simpson's rule is not the way  forward! 
Because the integrand is positive definite, the standard approach is to
use a Monte Carlo approximation  to the integrand.  This is implemented
in an `importance sampling' version  so that a stochastic estimate of
the integral is made from a finite  number of samples (called
configurations) of equal weight. The construction  of efficient
algorithms to achieve this is a topic in itself. Here I  will
concentrate on the analysis of the outcome, assuming that such 
configurations have been generated.  

So what is at our disposal is a set of samples of the vacuum. It  is
then straightforward to evaluate the average of various products  of
fields over these samples --- this gives the Green functions by
definition.  The Green function can then be continued from Euclidean to
Minkowski time (in most cases this  is trivial)  and compared to
experiment. Thus masses and matrix elements can be evaluated readily.
Scattering,  hadronic decays, real-time processes etc will not be
directly accessible.

 A simple example of a lattice measurement is that of a Wilson loop. 
This is the expectation value in the vacuum of the path  ordered product
of links around a  closed loop. For example, a rectangular loop   of
size $R \times T$ can be used to extract the potential between  heavy
quarks and thus to explore confinement.

 So far I have concentrated on the gluonic degrees of freedom of QCD,
where an elegant formulation was available. In contrast, the inclusion
of quarks  in a lattice approach is very inelegant and computationally
challenging. The  quarks will be represented by Grassmannian variables
$\psi$ and the fermionic  action for each species of quark will be
bilinear $\bar{\psi}Q(m,U) \psi$  where $Q$ is the fermionic matrix for
a quark of mass $m$ in the gauge fields $U(x)$, namely  $\not{\! \! D}+m$
in the continuum. Various discretisations of this have been proposed. 
They all have the feature that each fermionic species needs to be at
least doubled in a local lattice formulation. The Wilson fermionic
approach  kills the doublers at the expense of breaking chiral symmetry
while the staggered (Kogut-Susskind) approach  retains chiral symmetry
but mixes space, flavour and colour degrees of freedom. Both
discretisations  are expected to yield the same continuum result as $a
\to 0$.

 The treatment of these discretised fermions is still difficult. The
fermionic contribution to the action is not positive  definite so
straightforward Monte Carlo methods are excluded. The  usual approach is
to exploit the fact that the action is bilinear to explicitly integrate
out the $N_f$ Grassmannian fields  leaving an effective action in terms
of the gauge fields:
 \begin{equation}
   S = \sum S_{\Box} + N_f {\rm Tr} \log [ Q(m,U) ]
 \end{equation}
 Algorithms  to deal with the non-local trace-log term exist but they
are very  computationally intensive. It is frustrating that adding quark
degrees of freedom  results in a computational increase by  a factor of
1000 or more. This is the reason that the so-called {\em quenched}
approximation  is so popular. Here the limit $N_f=0$ is taken in
constructing vacuum samples --- i.e. just pure gluonic  QCD. Then quarks
can be propagated in this gluonic vacuum by solving  the lattice Dirac
equation ($Q(m,U) \psi = 0$). This approach is not unitary --- the
quarks do not feel any back reaction from quark pairs in the vacuum.
However,  it appears to be a rather good approximation for many
purposes. 

Going beyond the quenched approximation,  most approaches use 2 flavours
of equal mass quarks  to give a satisfactory algorithm and then vary the 
quark mass. The limit of large quark mass is of course just the quenched
approximation  since heavy quark loops are suppressed. 

The validation of the lattice approach calls for a series of checks 
that everything is under control.  
 \begin{itemize}  
 \item the lattice spacing should be small enough (discretisation 
errors) 
 \item the lattice must be big enough in space and time (finite size
errors)
 \item the statistical errors must be under control  
 \item Green functions must be extracted  with no contamination (eg a
ground state mass could be contaminated  with a piece coming from an
excited state) 
 \item both the quark contribution to the vacuum (sea quarks)  and the
quark constituents of hadrons (valence quarks) are  usually treated by
using larger mass values than the  experimental ones and then
extrapolating. This extrapolation must be treated accurately.
 \end{itemize}

The most subtle of these is the discretisation error. In order to 
extract the continuum limit of the lattice, one must show that the 
physical results will not change if the lattice spacing is decreased
further.  This is subtle because the lattice spacing is not known
directly --- in effect,  it is measured. I first discuss what is expected
on general  grounds to be the appropriate way to achieve small $a$. The
lattice simulation is undertaken at a value  of a parameter
conventionally called $\beta$ --- see eq(4). In the limit of small 
coupling $g^2$, where perturbation theory applies, $\beta=2N/g^2$. Thus 
large $\beta$ corresponds to small $g^2$. Now, perturbatively for $N=3$
colours,  the coupling $g^2$ corresponds to the  lattice spacing $a$  as
  \begin{equation}
g^2\approx {1 \over b_0 \log a^{-2}} \ \ \hbox{where}\ \  
b_0={11-{2 \over 3} N_f \over 16\pi^2} 
 \end{equation}
 for $N_f$ flavours of quarks. For the case of interest,
$N_f \le 3$, this corresponds to small values of $g^2$ at small distance
scale $a$ ---  as expected from  asymptotic freedom.

The perturbative argument is appropriate  to the study of results at
large  $\beta$ (small $g^2$). The lattice simulation  of QCD uses values
of $\beta \approx 6$ and hence bare couplings  corresponding to $\alpha
= g^2/(4\pi) \approx 0.08$. In the pioneering  years of lattice work,
this was thought to be a sufficiently small  number that the
perturbation series would converge rapidly. One of  the major advances,
in recent years, has been the realisation that  the bare lattice
coupling (our $g^2$ above) is a very poor  expansion parameter and the
perturbation series in the bare coupling does not converge well  at the
$\beta$ values of interest.  The theoretical explanation for this poor
convergence is that the  lattice action differs from the continuum
action and allows extra  interactions. These include tadpole
diagrams\site{LM} which have the property that  they sum up to  give a
contribution that involves high order terms in the  perturbation series.
The way to avoid this problem with tadpole terms is to  use a
perturbation series in terms of a renormalised coupling ---  rather than
the bare lattice coupling. I return to this topic  when discussing the
lattice determination of  the QCD coupling $\alpha$ later.

This change of attitude to the method of determining $a$ from  $\beta$
has had considerable implications  for lattice predictions, as I now
explain, since the aim is to work in a region of lattice spacing $a$ 
where perturbation  theory in the bare coupling is not precise. Because
of this,  in practice, $a$  is determined from the non-perturbative
lattice results themselves.  Thus if the energy of some particle is
measured on the  lattice, it will be available  as the dimensionless
combination $\hat{E}$. From a value for $E$ in physical units,  then $a$
can be determined since, on dimensional grounds, $\hat{E} =Ea$.
Furthermore, by increasing $\beta$, the  change in the observable
$\hat{E}$ gives information about the change in $a$  since $E$ is fixed
--- assuming it is the physical mass.  This should allow  a calibration
of $\beta$ in terms  of $a$ to be established.
 
This procedure is overly optimistic, however. The discretised lattice 
theory is different from the continuum theory on scales of the  order of
the lattice spacing $a$. For the Wilson action formulation  of gauge
theory, this implies that the continuum energy $E_c$ is related  to the
lattice observable $\hat{E}(\beta) = Ea$ as
  \begin{equation}
 { \hat{E} (\beta) \over a }  = E_c + {\cal O}(a^2) 
\end{equation}
 A direct consequence is that the ratio of two energies (of 
different particles, for example) will have  discretisation 
errors of order $a^2$. 

Thus, to cope with discretisation errors, the procedure required is 
to evaluate dimensionless ratios of quantities of physical interest 
at a range of values of the lattice spacing $a$ and then extrapolate 
the ratio to the continuum limit ($a \to 0$).

Note that when the fermionic terms are included, the discretisation 
error is of order $a$ for the Wilson fermionic action. By adding 
further terms in the fermionic action, the error can be reduced --- 
to order $\alpha a$ for the SW-clover fermion formulation.

\section{GLUEBALL MASSES}

I choose to illustrate the workings of the lattice method by describing 
the determination of the glueball spectrum.  Of course, glueballs are 
only defined unambiguously in the quenched approximation --- where 
quark loops in the vacuum are ignored. In this approximation, glueballs 
are stable and do not mix with quark-antiquark mesons.  This
approximation  is very easy to implement in lattice studies: the full
gluonic action  is used but no quark terms are included. This
corresponds to a full  non-perturbative treatment of the gluonic degrees
of freedom in  the vacuum. Such a treatment goes much further than
models such as the  bag model.

The glueball mass can be measured on a lattice through evaluating the
correlation $C(t)$ of two closed  colour loops (called Wilson loops) at
separation $t$ lattice  spacings. Formally 
  \begin{equation}  
 C(t) = \langle{}0|G(0) G^{\dag}(t)|0\rangle{} 
 = \sum_{i=0} c_i \langle{}g_i(0)|g_i(t)\rangle{} c^*_i 
 = \sum_{i=0} |c_i|^2 e^{-\hat{m}_i t } 
 \end{equation}  
 where $G$ represents the closed colour loop which can be thought  of as
creating a glueball state $g_i$ from the vacuum. Summing  over a
complete set of such glueball states (strictly these are eigenstates  of
the lattice transfer matrix  where $\exp{-\hat{m}_i}$ is the lattice
eigenvalue corresponding to a step of one lattice spacing in time) then
yields the above  expression. As $t \to \infty$, the lightest glueball 
mass will dominate.  This can be expressed as 
  \begin{equation}
\hat{m}_0 = \lim_{t \to \infty}\hat{m}_{\rm eff}(t) \ \
\hbox{where} \ \   \hat{m}_{\rm eff}(t) = \log({ C(t-1) \over C(t)})
 \end{equation}
Note that since for the excited states $\hat{m}_i > \hat{m}_0$, then 
$\hat{m}_{\rm eff}(t) > \hat{m}_{\rm eff}(t+1) > \hat{m}_0$. This 
implies that the effective mass, defined above, is an upper bound  on
the ground state mass.  In practice, sophisticated methods are used to
choose loops $G$ such  that the correlation $C(t)$ is dominated by the
ground state glueball  (i.e. to ensure $|c_0| >> |c_i|$). By using 
several different loops,  a variational method can be used to achieve
this effectively.  These techniques are needed to obtain accurate
estimates of $\hat{m}_0$ from  modest values of $t$ since the signal to
noise decreases as $t$ is  increased. Even so, it is worth keeping in
mind that upper  limits on the ground state mass are obtained in
principle.

The method also needs to be tuned to take account of the many 
glueballs: with different $J^{PC}$ and different momenta. On the lattice
the Lorentz symmetry is reduced to that of  a hypercube. Non-zero 
momentum sates can be created (momentum is discrete in units of  $2 \pi
/L$ where $L$ is the lattice spatial size). The usual  relationship
between energy and momentum is found for sufficiently  small lattice
spacing. Here I shall concentrate on the simplest  case of zero
momentum (obtained by summing the correlations over  the whole spatial
volume). Charge conjugation $C$ is a good  quantum number in lattice
studies of glueballs: $C$ interchanges the direction of  all links.  

For a state at rest, the rotational symmetry becomes a  cubic symmetry.
The lattice states (the $g_i$ above) will transform  under irreducible
representations of this cubic symmetry group (called $O_h$). These
irreducible representations can be linked to  the representations of the
full rotation group SU(2). Thus, for  example, the five spin components
of a $J^{PC}=2^{++}$ state should be appear as  the two-dimensional 
E$^{++}$ and the three-dimensional T$_2^{++}$ representations on  the
lattice, with degenerate masses. This degeneracy requirement then
provides a  test for the restoration of rotational invariance ---  which
is expected to occur at sufficiently small lattice spacing.

The results of lattice measurements\site{DForc,MT,glue,gf11} of the
$0^{++}$ and  $2^{++}$ states are shown in fig.~1. The restoration of
rotational  invariance is shown by the degeneracy of the two $O_h$
representations that make up the $2^{++}$ state.  Fig.~1 shows the
dimensionless combination of the  lattice glueball mass $\hat{m}_0$ to a
lattice quantity $\hat{r}_0$. I will return to describe  the lattice
determination of $\hat{r}_0$ in more detail --- here it suffices to accept
it as a well measured quantity on the lattice that can be used to
calibrate the lattice spacing and so explore the continuum limit. The
quantity plotted, $ \hat{m}_0 \hat{r}_0$, is expected to be equal to the
product of continuum quantities $m_0 r_0$ up to corrections of order
$a^2$. This is indeed  seen to be the case. The extrapolation to the
continuum limit ($a \to 0$) can then be made with confidence.
This is an important result: continuum quenched QCD does have  massive 
states and their properties are determined.

 The only other candidate for a relatively light glueball is the
pseudoscalar. Values quoted of $\hat{r_0} \hat{m(0^{-+})}= 5.6(6),\
7.1(1.1)$ and 5.3(6) from refs(\cite{MT,glue}) suggest  an average of
6.0(1.0), not appreciably lighter  than the tensor glueball. This is
confirmed by preliminary results from the group  of ref(\cite{mpglue})
that the pseudoscalar is heavier than the tensor glueball.

The value of $r_0$ in physical units is about 0.5 fm and I will  adopt a
scale equivalent to $r_0^{-1} = 0.372 $ GeV with a 10\% systematic error
on this scale since different physical observables differ from the 
quenched approximation values by this amount.  Then the  masses of
lightest glueballs in the continuum limit are $m(0^{++})=1611(30)(160)$
MeV;  $m(2^{++})=2232(220)(220)$ MeV and $m(0^{-+})=2232(370)(220)$ MeV,
where the second error is the overall scale error.

Recently a lattice approach using a large spatial lattice spacing with
an  improved action and a small time spacing has been used to study
glueball  masses. The results~\site{mpglue} in the continuum limit are
that $r_0m(0^{++})=3.98(15)$,  $r_0m(2^{++})=5.85(2)$,
$r_0m(1^{+-})=7.21(2)$ and $r_0 m'(2^{++})=8.11(4)$. There remains a
small discrepancy with the result for the $0^{++}$ glueball obtained
above ($r_0m(0^{++})=4.33(5)$) from lattice spacings much closer to the
continuum limit. When this is fully  understood, the new method looks to
be very promising for  access to excited glueball masses.

\begin{figure}[ht]
\vspace{13.5cm} 
\includegraphics{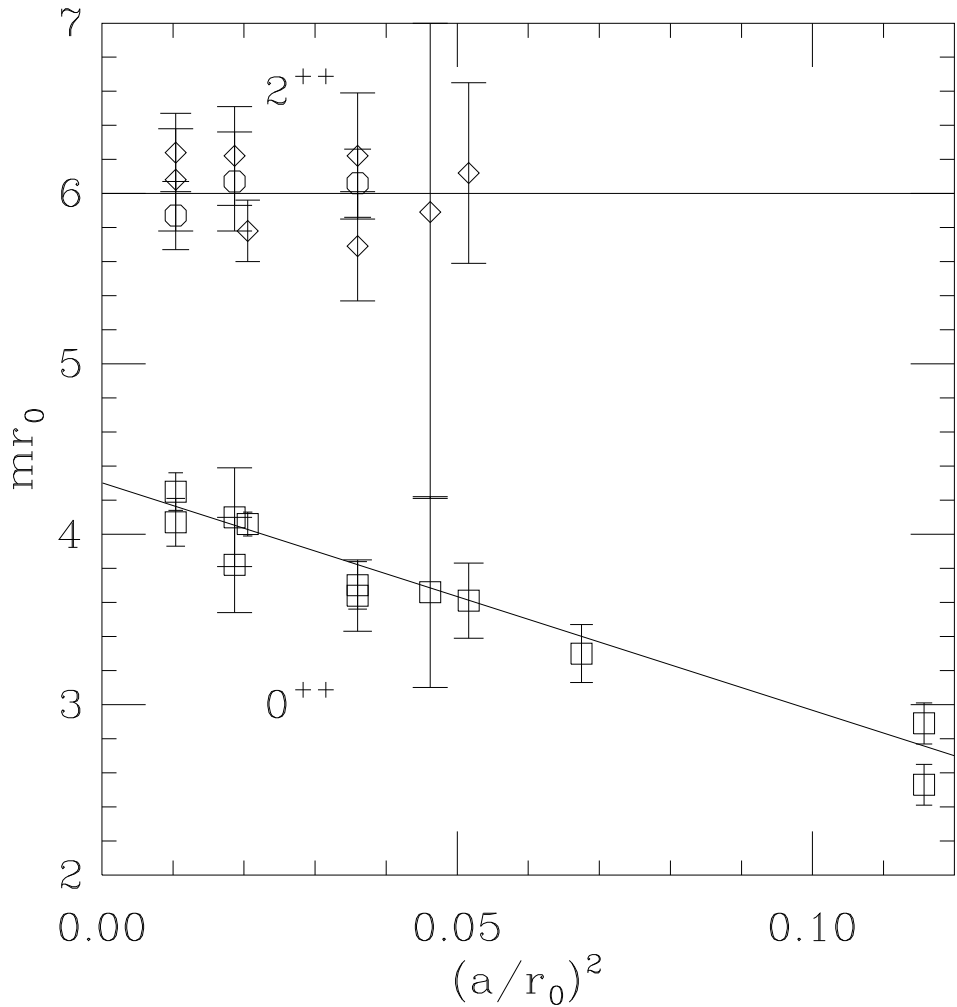}
 \caption{ The value of mass of the  $J^{PC}=0^{++}$ and $2^{++}$
glueball states from refs({\protect\cite{DForc,MT,glue,gf11}}) in units
of $r_0$. The $T_2$ and $E$ representations are shown  by  octagons  and
diamonds respectively.  The straight line  shows a fit describing the
approach to the continuum limit as $a \to 0$.
   }
\end{figure}

\begin{figure}[ht]
\vspace{16cm}  
\includegraphics{gbspect.fig}
  \caption{ The  mass of the  glueball states  with quantum  numbers
$J^{PC}$ from ref({\protect\cite{glue}}).  The scale is set by
${\protect\sqrt{\sigma}} \approx 0.44 $ GeV which yields the right hand
scale in GeV. The solid points represent mass determinations whereas the
open points  are upper limits.
   }
\end{figure} 

The predictions for the other $J^{PC}$ states are that they lie 
higher in mass and the present state of knowledge is summarised 
in fig.~2.  Remember that the lattice results are strictly 
upper limits. For the   $J^{PC}$ values not shown, these 
upper limits are too weak to be of use.

Since quark-antiquark mesons can only have certain  $J^{PC}$ values, 
it is of special interest to look for glueballs with  $J^{PC}$ values 
not allowed for such mesons: $0^{--}$, $0^{+-}$, $1^{-+}$, $2^{+-}$,
etc.  Such spin-exotic states, often called `oddballs',  would not mix
directly with quark-antiquark mesons. This would  make them a very
clear experimental signal of the underlying glue dynamics.  Various 
glueball models (bag models, flux tube models, QCD sum-rule  inspired
models,..) gave different predictions for the presence of such oddballs
(eg. $1^{-+}$) at relatively low masses. The lattice mass spectra
clarify these uncertainties but, unfortunately for experimentalists,  do
not indicate any low-lying oddball candidates. The lightest candidate 
is from the T$_2^{+-}$ spin combination. Such a state could correspond 
to an $2^{+-}$ oddball. Another interpretation is also possible,
however,  namely that a non-exotic $3^{+-}$ state is responsible (this
choice of interpretation can be resolved in principle by finding the 
degenerate 5 or 7 states of a $J=2$ or 3 meson). The overall  conclusion
at present is that there is no evidence for any oddballs  of mass less
than 3 GeV.

Returning briefly to the independence of the results on the volume  of
the lattice, in the early days of glueball mass determination, it  was
expected that a spatial size $L$ should satisfy  $\hat{m}(0^{++})L > 1 $
and, hence, that values of $\hat{m}(0^{++})L $ of 1 to 4 would suffice.
A careful lattice  study\site{smallg}  showed that $\hat{m}(0^{++})L >
8 $ was required to  obtain rotational invariance and a result
independent of $L$. The results  collected in fig.~1 all satisfy this
latter inequality so can be regarded as the infinite volume
determination. 

The small volume lattice results illustrate several points. One
important  feature is that closed loops of colour flux acting on the
vacuum   create glueballs but, if the loop encircles the periodic
spatial boundary of length $L$,  a torelon state can be created with
energy given approximately by $KL$ where $K$ is the string tension. Thus
at small $L$,  the torelon state will be lighter than a glueball. In the
quenched approximation, there is  a $Z(N)$ symmetry of the lattice
action which puts glueballs and torelons in different representations 
so they are unmixed. The fermion term in the action, however, breaks
this symmetry  so that for $N_f \not = 0$, the lightest glueball will be
very heavily  contaminated by torelon-like contributions. In a
semi-analytic study\site{krip}  this was explored and what is clear is
that  in  a small volume the limit $N_f \to 0$ of full  QCD is {\em not}
equivalent to the quenched approximation with $N_f=0$.

Glueballs are defined in the quenched approximation and, hence, they  do
not decay into mesons since that would require quark-antiquark 
creation. It is, nevertheless, still possible to estimate the  strength
of the matrix element between a glueball and a pair of mesons  within
the quenched approximation. For the glueball to be a relatively narrow 
state, this matrix element must  be  small. A  very preliminary attempt
has been made to estimate the size of  the coupling of the $0^{++}$
glueball to two pseudoscalar mesons\site{gdecay}. A relatively small
value of order 100 MeV is found. Further work needs to be done to 
investigate this in more detail, in particular to study the mixing
between  the glueball and $0^{++}$ mesons since this mixing may be an
important  factor in the decay process.

Another lattice study  will become feasible soon. This is to study  the
glueball spectrum in full QCD vacua with sea quarks of mass $m_{\rm
sea}$. For large $m_{\rm sea}$, the result is just the quenched result
described above.  For $m_{\rm sea}$ equal to the experimental light
quark masses, the results  should just reproduce the experimental meson
spectrum --- with the  resultant uncertainty between glueball
interpretations and other  interpretations. The lattice enables these
uncertainties to be resolved in principle: one obtains the spectrum for
a range of values of $m_{\rm sea}$  between these limiting cases, so
mapping glueball states at large  $m_{\rm sea}$ to the experimental
spectrum at light $m_{\rm sea}$.  Studies  conducted so far show no
significant change of the glueball spectrum as dynamical quark effects
are added --- but  the sea quark masses used are still rather
large\site{sesam}.

 From the point of view of comparing quenched lattice results with
models,  a very useful system to study is the {\em gluelump}. This is
the  hydrogen atom of gluonic QCD: a system with one very heavy gluon
which is  treated as a static adjoint colour source and the surrounding
gluonic field that  makes a colour singlet hadron. In terms of
experiment: this is the gluinoball formed from a gluino-gluon bound
state which will be observable should  massive gluinos exist and be
sufficiently stable. The spectrum and spatial distribution of these
states  have been explored\site{cmjor} for $SU(2)$ of colour. Because
one  colour source is fixed, the spatial size is easier to measure than
for the  glueballs themselves: the distributions were found to extend
out to  a radius of 0.5 fm. The ground state and first excited state
were found to have quantum numbers consistent with being  bound states
of a magnetic gluon and an electric gluon respectively. For $SU(3)$ of
colour the spectrum  in the continuum limit has been
determined\site{cmmsf} and the mass splitting between the two lightest
of these  states  is found to be 350 MeV.

\section{POTENTIALS BETWEEN QUARKS}

A very straightforward quantity to determine from lattice 
simulation is the interquark potential in the limit of very 
heavy quarks (static limit). This potential is of direct 
physical interest because solving the Schr\"odinger equation in 
such a potential provides  a good approximation to the $\Upsilon$ 
spectrum.  It is also relevant to exploring both confinement 
and asymptotic freedom on a lattice.

The basic route to the static potential is to evaluate the average 
$W(\hat{R},\hat{T})$ in the vacuum samples of a rectangular closed loop
of colour flux  (a Wilson loop of size $\hat{R} \times \hat{T}$).  This
can be  visualised as involving static sources $R$ apart with potential
energy $V(R)$  for time $T$ so that $W \approx c e^{-V(R) T}$. More
precisely,  the lattice quantities $\hat{R}$ and $\hat{T}$ are related
to  the physical distances $R$ and $T$ by $\hat{R}=R/a$, etc where  $a$
is the lattice spacing which is not known explicitly. Then it can be
shown that the required static potential in lattice units is given by 
  \begin{equation} 
\hat{V}(\hat{R}) = \lim_{\hat{T} \to \infty} \log{W(\hat{R},\hat{T}-1)
\over W(\hat{R},\hat{T})} 
 \end{equation} 
 The limit of large $T$ is needed to separate the required potential
from  excited potentials. This limit can be made tractable in practice
by  using more complicated loops than the simple rectangular loop
described  above. The motivation for this  is to generalise the straight
paths of length $R$ at $t=0$ and $T$ by considering sums over paths that
reflect more fully the colour flux between static quark and antiquark
at  separation $R$. Typically a smearing or fuzzing algorithm is used 
to create suitable wandering paths, then several such paths are 
combined in a variational approach to find the linear combination that
best  describes the ground state of the system: the potential $V(R)$.

\begin{figure}[b!] 
\vspace{-2cm}
\epsfysize=13cm
\epsfig{
   file=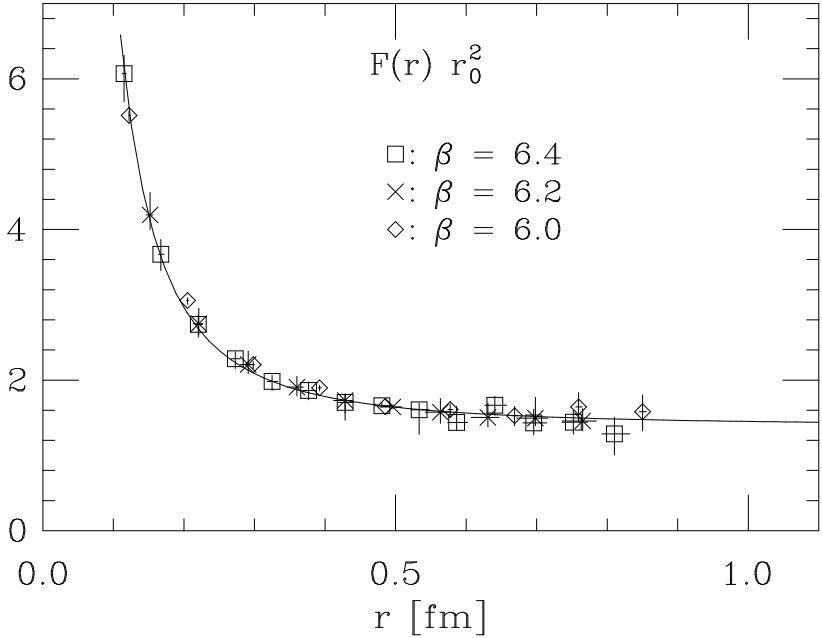, height=13cm}
  \caption{ The force between static quarks at separation $r$
as measured on lattices with different lattice spacing $a$ (with 
the $\beta$-values shown).
   }
\end{figure}

A summary of results\site{pots,cmper} for the potential at large $R$ is
shown in figs~3,\ 4.  The result that the force $dV/dR$ tends to a 
constant at large $R$ (and thus $V(R)$ continues to rise as $R$
increases) is  a manifestation of the confinement of heavy quarks (in
the quenched approximation). The force  appears to approach a constant
at large $R$. A simple parametrisation  is traditional in this field:
  \begin{equation}
V(R) = V_0 - { e \over R} + K R
 \end{equation}
where $K$ (sometimes written $\sigma$) is the string tension. The 
term $e/R$ is referred to as the Coulombic part in analogy to the 
electromagnetic case. The equivalent relationship in terms of 
quantities defined on a lattice is
  \begin{equation}
\hat{V}(\hat{R}) = \hat{V}_0 - { e \over \hat{R}} + \hat{K} \hat{R}
 \end{equation}.

\begin{figure}[ht!] 
\vspace{11cm} 
\includegraphics{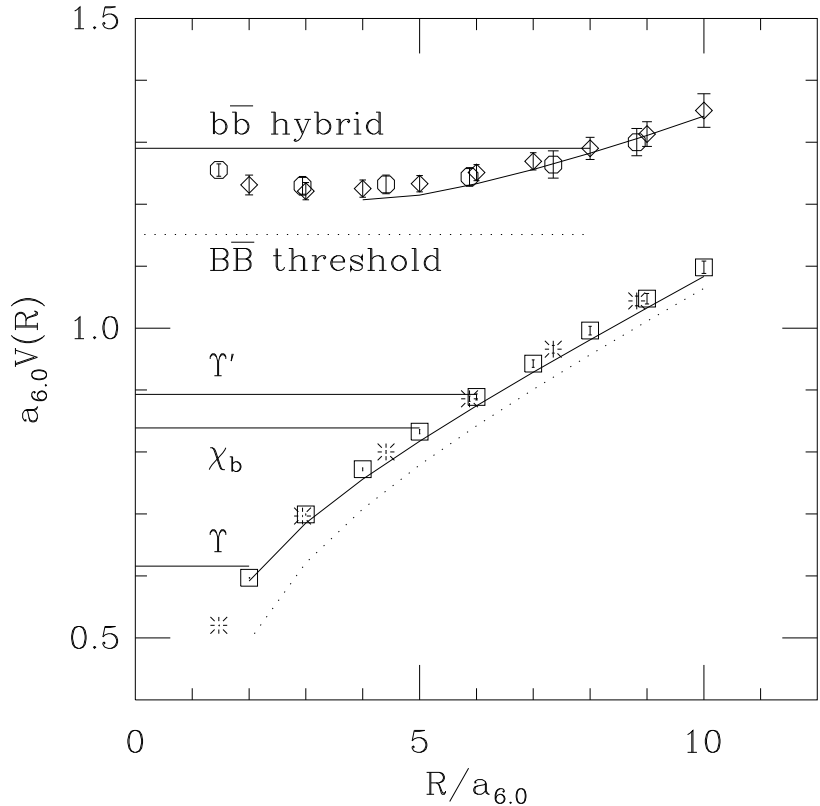}
 \caption{ Potentials $V(R)$ between static quarks  at separation $R$
for the  ground state ($\Box$ and {\tt *}) and for the $E_u$ symmetry
which corresponds to the first excited state of the gluonic flux
(octagons and diamonds). Results in lattice units ($a_{6.0}^{-1}=2.02$
GeV) from  the quenched calculations of ref({\protect\cite{cmper}}) are
shown by symbols corresponding to different lattice spacings. For the
ground state  potential the continuous curve is an interpolation of the
lattice data while the dotted  curve with enhanced Coulomb term fits the
spectrum and yields the masses shown. The lightest hybrid level in the
excited gluonic potential is  also shown.
   }
\end{figure}

Since the string tension is given by the slope of $V(R)$ against  $R$ as
$R \to \infty$, this implies that some error will arise in determining
$K$  coming from the  extrapolation of lattice data at finite $R$. A
practical resolution\site{sommer0}  is to define a value of $R$ where
the potential takes a certain form. The convention is to use $r_0$ where
  \begin{equation}
 \left. {\hat{R}^2 {d\hat{V}(\hat{R}) \over d\hat{R}}}
\right|_{\hat{r}_0} = 1.65
 \end{equation}
 Thus $\hat{r}_0$ can be determined by interpolation in $\hat{R}$ rather
than  extrapolation.  In practice, this means that $\hat{r}_0$ is very
accurately  determined by lattice measurements and so is a useful
quantity to use  to set the scale since $\hat{r}_0=r_0/a$. With the
simple parametrisation above,  $r_0^2=(1.65-e)/K$ so $r_0$ is
closely related to the string tension  since $e \approx 0.25$.  The
string tension is usually taken from  experiment as $\sqrt{K}=0.44 $GeV
where the value comes from $c \bar{c}$ and $b \bar{b}$ spectroscopy and
from the light  meson spectrum interpreted as excitations of a
relativistic string. Similar analyses also imply that $r_0 \approx 0.5 $
fm. Here I use $r_0^{-1}=0.372$ GeV  to be specific.  Since I shall be
describing quenched lattice results, the energy scale set from different
physical quantities will not  necessarily agree (since {\it experiment}
has full QCD not the quenched vacuum) and so a systematic error of order
10\% must be applied to any such choice of scale. This was discussed
when taking glueball mass  values from quenched lattice calculations.

The lattice potential $V(R)$ can be used to determine the spectrum  of
$b \bar{b}$ mesons by solving Schr\"odinger's equation since the  motion
is reasonably approximated as non-relativistic. The lattice result is
similar to the experimental $\Upsilon$ spectrum. The main  difference is
that the Coulombic part ($e$) is effectively too  small (0.28 rather
than 0.4). This produces\site{cmper} a ratio of  mass differences
$(1P-1S)/(2S-1S)$ of 0.71 to be compared  with the experimental ratio of
0.78.  This difference is understandable  as a consequence of the
Coulombic force at short distances which would be  increased by
$33/(33-2N_f)$ in perturbation theory in full QCD compared  to quenched
QCD. I will return to discuss this in the context of lattice dynamical 
fermion calculations\site{sesamv}.

\subsection{Running coupling constant}

At small $R$, the static potential can be used, in principle,  to study
the running coupling constant. Small $R$ corresponds  to large momentum
and  the coupling should decrease at  small $R$. Thus the Coulombic
coefficient $e$ introduced above should actually  decrease
logarithmically as $R$ decreases. Perturbation theory can  be used to
determine this behaviour of the potential at small $R$.

In the continuum the potential between static quarks is known
perturbatively to two loops in terms of  the  scale $\Lambda_{\msbar} $.
For  $SU(3)$ colour, the continuum force is given for $N_f=0$
by\site{bill}

\begin{equation}
{dV \over dR } =  {4 \over 3} {\alpha(R) \over R^2}
\end{equation}

\noindent with the effective coupling $\alpha (R)$ given by

\begin{equation}
 { 1 \over 4 \pi [ b_0  \log (R\Lambda _F )^{-2} + 
(b_1 / b_0 ) \log \log (R\Lambda_F )^{-2} ] }
\end{equation}

 \noindent where $b_0=11/16 \pi ^2$ and $b_1=102 \ b_0^2/121$ are the
usual coefficients in the  perturbative expression for the
$\beta$-function, neglecting quark loops in the vacuum.
 Here  $\Lambda_F= 1.048 \Lambda_{\msbar}$.

This perturbative result can be used to {\em define} a running coupling 
constant non-perturbatively. Thus a coupling in the `force' scheme 
can be defined by 
 \begin{equation}
\alpha_F(R) \equiv  {3 \over 4} R^2 {dV \over dR }   
 \end{equation}
 Different non-perturbative definitions of $\alpha$ can be related using
perturbation  theory, for example the one loop comparison of $\alpha_F$
and $\alpha_{\msbar}$ gives  the relationship between the respective
$\Lambda$ values quoted above.

On a lattice the force can be estimated by  a finite difference and one
can  extract  the  running  coupling constant by using\site{cmlms}

\begin{equation} 
 \alpha_F( { \hat{R}_1 + \hat{R}_2 \over 2 }) = { 3\over 4} \hat{R}_1
\hat{R}_2 { \hat{V}(\hat{R}_1)-\hat{V}(\hat{R}_2) \over 
\hat{R}_1-\hat{R}_2 } 
 \end{equation}

\noindent where the error in using a finite difference is  negligible in
practice.

This  is plotted  in  fig.~5 versus $R  K^{1/2}$: this combination  is
dimensionless and  so can be  determined from lattice results since $R
 K^{1/2}=\hat{R} \hat{K}^{1/2}$,  where $\hat{K}$ is taken from the
fit to $\hat{V}(\hat{R})$. The interpretation of $\alpha_F$ as defined
above as an effective running coupling constant is only justified at
small $R$ where the  perturbative expression dominates. Also shown  are 
the two-loop perturbative results for $\alpha(R)$ for  different values
of $\Lambda_F $.

\begin{figure}[ht]
\vspace{12cm} 
\includegraphics{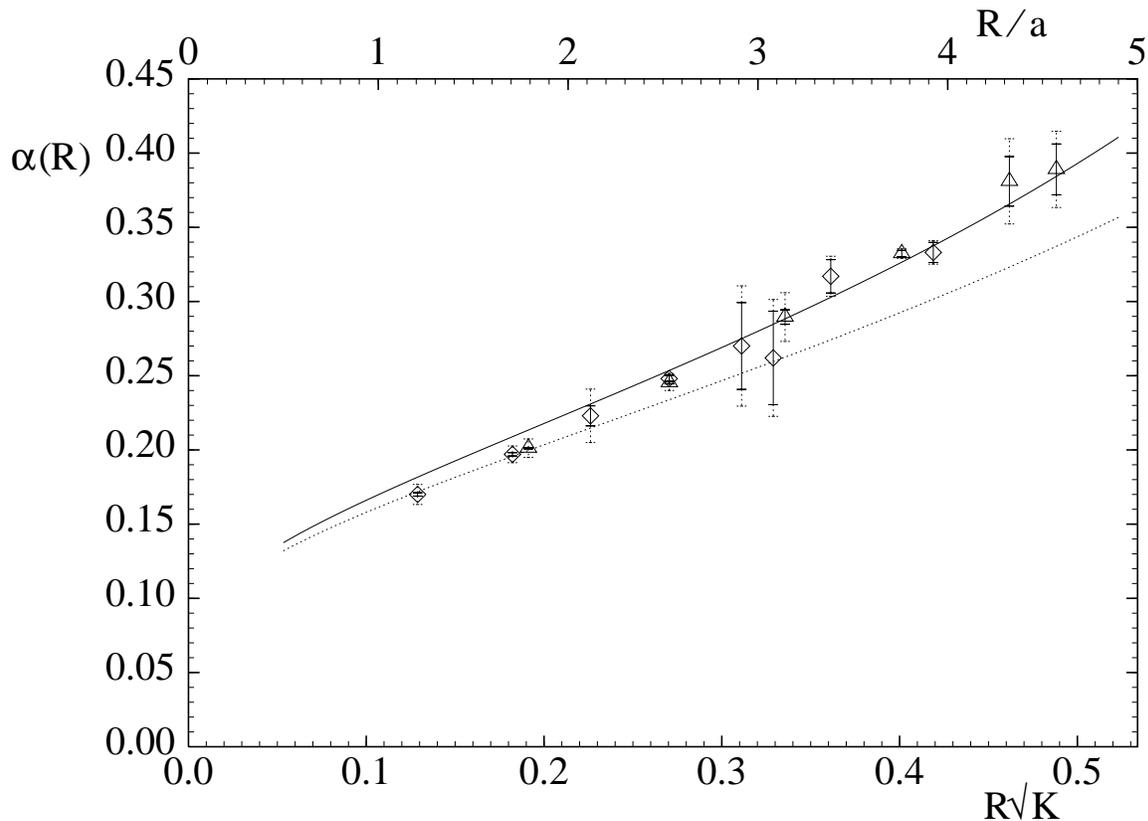} 
\caption{
 The effective running coupling constant $\alpha_F(R)$ obtained from  the
force between static quarks at separation $R$ from
ref({\protect\cite{uksu3}}).  The scale is set by the string tension $K$.
Data are at $\beta=6.5$ (diamonds) and at $\beta=6.2$ (triangles).  The
dotted error bars represent an estimate of the systematic error due to
lattice artefact corrections.  The curves are the two-loop perturbative
expression with $a(6.5)\Lambda_F=0.060$ (dotted) and 0.070 (continuous).
 }
\end{figure}

Fig.~5 clearly shows a {\it running} coupling constant.  Moreover
the result is consistent with the expected perturbative dependence on
$R$ at small $R$.  There are systematic errors, however. At larger $R$,
the perturbative two-loop expression will not be an accurate estimate of
the measured potentials, while at smaller $R$, the lattice artefact
corrections (which arise because $R/a \approx 1$) are relatively big. 
Setting the scale using $\sqrt K=0.44$ GeV implies $1/a(\beta=6.5)=4.13
$ GeV, so $R < 4a(6.5)$ corresponds to values of $1/R > 1$ GeV.  This
$R$-region is expected to be adequately described by perturbation
theory. Other methods to extract a running coupling  from lattice
results have been used and some have smaller systematic errors.  For a
comparison see ref({\cite{cmlat94}).

This determination from the interquark force of the coupling $\alpha$ 
allows us to compare the result with the bare lattice coupling
determined  from $\beta=6/g^2$. At $\beta=6.5$, $\alpha_{\rm
bare}=g^2/4\pi=0.073$.  The values of $\alpha$ shown in fig.~5 are much
larger. The effective  coupling constant is thus almost twice the bare
coupling. This is  quite acceptable in a renormalisable field theory.
The message is that  the bare coupling should be disregarded --- it is
not a good expansion  parameter. The measured $\alpha_F$, however, proves
to be a reasonable  expansion parameter in the sense that  the first few
terms of the perturbation series converge.  This successful calibration
of  perturbation theory on a lattice is important in practice. For
instance,  when matrix elements are measured on a lattice they have
finite correction factors (usually called Z) to relate them to continuum
matrix elements. These  Z factors are evaluated perturbatively --- so an
accurate continuum  prediction needs trustworthy perturbative
calculations.

\subsection{Excited gluonic modes}

The situation of a static quark and antiquark is a very clear case in
which to discuss hybrid  mesons which  have excited gluonic
contributions. A discussion of the colour  representation of the quark
and antiquark is not useful since they are at different space positions
and the combined colour is not gauge invariant. A better criterion is 
to focus on the spatial symmetry of the gluonic flux. As well as the
symmetric ground state of the colour flux between two  static quarks,
there will be excited states with different symmetries.  These were
studied on a lattice\site{liv} and the conclusion was that the $E_u$
symmetry (corresponding to flux  states from an  operator which is the
difference of U-shaped paths from quark to antiquark of the form $\,
\sqcap - \sqcup$) was the lowest lying gluonic excitation.  Results  for
this potential are shown in fig.~4.

This gluonic excitation corresponds to a component of angular momentum
of one unit along the quark antiquark axis. Then one can solve for the
spectrum of hybrid mesons using the Schr\"odinger equation in the
adiabatic approximation. The spatial wave function necessarily has non
zero angular momentum and  corresponds to $L^{PC}=1^{+-}$ and $1^{-+}$.
Combining with the  quark and antiquark spins then yields\site{liv} a
set of 8 degenerate hybrid states with $J^{PC}=1^{--},\ 0^{-+},\
1^{-+},\ 2^{-+}$ and    $1^{++},\ 0^{+-},\ 1^{+-},\ 2^{+-}$ 
respectively. These contain the  spin-exotic states with $J^{PC}= 
1^{-+},\ 0^{+-}$ and $2^{+-}$ which will be of special interest. 

 Since the lattice calculation of the ground state and hybrid masses
allows  a direct prediction for their difference, the result for this
8-fold degenerate hybrid level is illustrated in fig.~4  and
corresponds\site{cmper} to masses of 10.81(25) GeV for $b\bar{b}$ and 
4.19(15) GeV for $c \bar{c}$. Here the errors take into account the
uncertainty in setting the ground state mass using the quenched
potential as discussed above. Recently a different lattice
technique\site{morn} has been used to  explore the excited gluonic
levels in the quenched approximation. The results above are confirmed
and preliminary values quoted for the lightest hybrid mesons are 10.83
and 4.25 GeV respectively for $b\bar{b}$ and $c \bar{c}$ with no error
estimates given.

 The quenched lattice results show that the lightest hybrid mesons lie 
above the open $B \bar{B}$ threshold and are likely to be relatively
wide  resonances. This could also be checked by  comparing with quenched
masses for the $B$ meson itself\site{sommer}, but at present there are
quite  large uncertainties on that mass determination. The very flat
potential implies a very extended wavefunction: this has the implication
that the wavefunction at the origin will be small,  so hybrid vector
states will be weakly produced from $ e^+ e^-$.

 It would be useful to explore the splitting among the 8 degenerate
$J^{PC}$ values obtained. This could come from different excitation 
energies in the $L^{PC}=1^{+-}$ (magnetic) and $1^{-+}$
(pseudo-electric) gluonic excitations, spin-orbit terms, as well as
mixing between hybrid states and $Q\bar{Q}$ mesons with non-exotic spin.
One way to study this on a lattice is to use the  NRQCD formulation
which describes non-static heavy quarks which propagate 
non-relativistically. Preliminary results for hybrid excitations from
several  groups\site{alik} give at present similar
results to those with the static approximation as described above,  but
future results may be more precise and able to measure splittings among
different states.

 As well as comparing excited gluonic states from the lattice with
experimental spectra  for $\bar{b} b$ systems, it is also worthwhile to
compare with phenomenological models.  One such model is the hadronic
string. This has the simple prediction that  gluonic excitations are at
multiples of $\pi/R$ in energy higher at large $R$. A detailed 
comparison\site{cmper} for SU(2) of colour shows qualitative agreement
of  the lattice excited potentials with the  string model provided an
appropriate expression is used for excited level $j$:
 \begin{equation}
  V_j(R) = (K^2 R^2 - {\pi K \over 6} +2 \pi j K)^{1 \over 2}.
 \end{equation}
 This expression also shows that even the ground state string mode
($j=0$) will have  a contribution from a string fluctuation,
namely\site{lu} $V(R) \approx KR -\pi/(12R) $ as $R \to \infty$. In
practice this  $1/R$ string fluctuation term is very hard to disentangle
from the  Coulomb term $e/R$. One way to get round this in lattice
studies is to consider a hadronic string that  encircles the periodic
spatial boundaries of length $L$. Then there are no  sources and hence
no Coulomb component. The appropriate string fluctuation term in the 
energy of this state, called the torelon, is then given by  $E(L)=KL +
\pi/(3L)$ as $L \to \infty$. Lattice studies\site{cmpms}  have confirmed
the presence of this string fluctuation term with the correct
coefficient as given  to a precision of 3\%. This is impressive evidence
that the hadronic string is a good model  of the energy of the colour
flux tube at large distances. 

\subsection{Confinement }

   The simplest manifestation of confinement is that the  potential
$V(R)$ between static colour sources in the fundamental representation
continues to rise with increasing $R$ in quenched QCD. This raises  the
question of the nature of the colour fields between the sources. 
Lattice studies have been undertaken\site{wupp,hel} to probe the energy
momentum tensor of these  colour fields.  The probe used is a plaquette 
so the momentum scale of the probe increases as $a \to 0$. Lattice sum 
rules~\site{hel} can be used to normalise these distributions and relate
them to  the $\beta$-function.  For separation $R > 0.7$ fm, a
string-like spatial  distribution is found where the transverse width of
the flux tube increases slowly at most with  increasing $R$. For these
$R$ values, the averages of the squared components of the colour  fields
(these are gauge invariant quantities) are found to be roughly  equal
(i.e. $E^2_i \approx B^2_j$, for all $i,\ j = 1,\ 2,\ 3$). This implies
that  the energy density is much smaller than the action density.  As
well as components of the energy-momentum tensor which are  gauge
invariant, it is possible to study the colour field tensors  directly by
choosing a suitable gauge\site{pisa}.

  Another detail concerning the confining force is its spin dependence. 
A lattice study of the spin-spin and spin-orbit potentials between
static quarks  allows this to be explored. The basic
conclusion\site{cmLS} is that the only  long-range force is a
spin-orbit force of the type usually called scalar.  More recent
studies\site{wuppss} confirm this. This lattice result  is confirmed by
phenomenological studies of the observed  splitting between P-wave
$\bar{b}b$ and $\bar{c}c$ states.

 Another route to explore confinement is to measure on a lattice the
potential energy between  two static sources in the adjoint
representation of the colour group. At large $R$ the adjoint potential
$V_A(R)$  must become a constant because each adjoint colour source can
be screened by a gluonic field. Indeed at large $R$, $V_A \to 2 m_{\rm
gluelump}$ where the  gluelump is  the ground state hadron with a gluon
field around a static adjoint: the  gluinoball. Of interest to model
builders is the adjoint potential at  smaller $R$ values. The most
precise data are for $SU(2)$ of colour\site{cmadj} and they do show a
region of linear rise, although with a slope less steep than  that given
by the Casimir ratio (namely ${8 \over 3} V(R)$ where  $V(R)$ is 
the fundamental colour source potential discussed previously). Attempts
have also been made\site{trot} to explore the colour field 
distributions in this case.

\section{LIGHT QUARKS --- HYBRIDS}

 Unlike very heavy quarks, light quark propagation in the gluonic vacuum
sample is very computationally intensive --- involving inversion of huge
($10^7 \times 10^7$) sparse matrices. Current computer power is 
sufficient to study light quark physics thoroughly in the quenched 
approximation. The state of the art\site{yoshie} is the Japanese CPPACS
Collaboration  who are able to study a range of large lattices (up to
about $64^4$) with a range of light quark masses. Qualitatively the 
meson and baryon spectrum of states made of  light and strange quarks is 
reproduced with discrepancies of order 10\% in the quenched approximation.

\begin{figure}[bt!] 
\vspace{11cm} 
\includegraphics{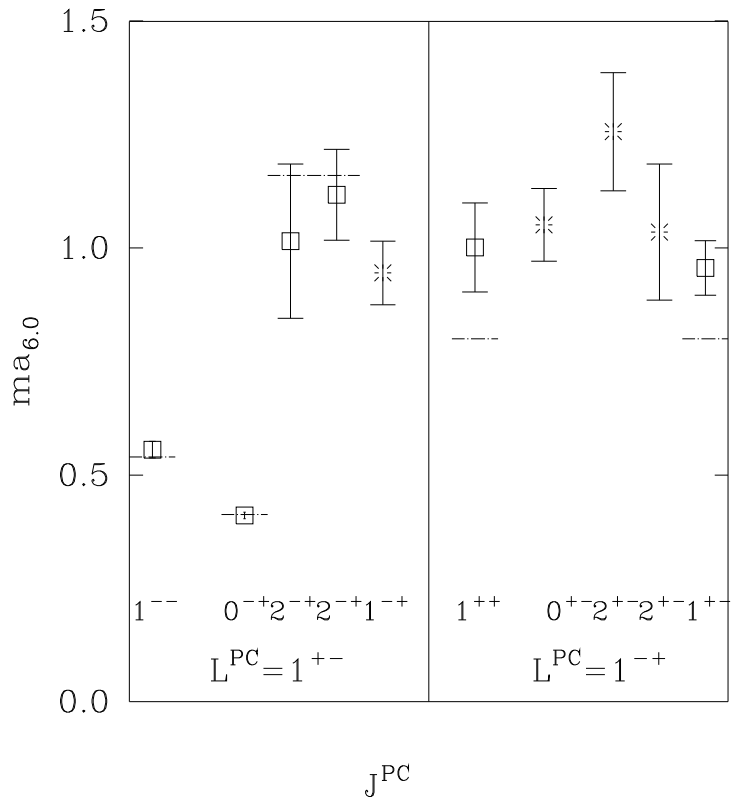}
 \caption{ The masses in lattice units (with $a_{6.0}^{-1} \approx 2$
GeV) of states of $J^{PC}$ built from hybrid operators with strange
quarks, spin-exotic ({\tt *}) and non-exotic ($\Box$). The dot-dashed
lines are the mass values found for  $s\bar{s}$ operators. Results from 
 ref({\protect\cite{hybrid}}). 
   }
\end{figure}

 Here I will focus on hybrid mesons made from light quarks. In the
quenched  approximation, there will be no mixing involving spin-exotic
hybrid mesons  and so these are of special interest. The first study of
this area was by the  UKQCD Collaboration\site{hybrid} who used
operators motivated by the  $Q\bar{Q}$ studies referred to above. Using
non-local operators, they studied  all 8 $J^{PC}$ values coming from
$L^{PC}=1^{+-}$ and $1^{-+}$ excitations. The  resulting mass spectrum
is shown in fig.~6 where the $J^{PC}=1^{-+}$ state  is seen to be the
lightest spin-exotic state with a statistical significance of 1 standard
deviation. The statistical error on the mass of this lightest
spin-exotic meson  is 7\% but to take account of systematic errors from
the lattice determination a  mass of 2000(200) MeV is quoted for the $s
\bar{s}$ meson. Although not  directly measured, the corresponding light
quark meson would be expected to be around 120 MeV lighter. In view of
the   small statistical error, it seems unlikely that the $1^{-+}$ meson
in the quenched approximation could lie as light as 1.4 GeV where there
are experimental  indications for such a state\site{hadron97}. Beyond
the quenched  approximation, there will be mixing between such a hybrid
meson and $q \bar{q} q \bar{q}$ states such as $\eta \pi$ and this may
be significant.

One feature clearly seen in fig.~6 is that non spin-exotic mesons
created  by hybrid meson operators have  masses  which are very similar
to those found when the states are created by $q \bar{q}$ operators.
This suggests that there is  quite strong coupling between hybrid and $q
\bar{q}$ mesons even in the quenched approximation. This would imply that 
the $\pi(1800)$ is unlikely to be a pure hybrid, for example.

A second lattice group has also evaluated hybrid meson spectra from
light quarks. They obtain\site{milc} masses with statistical and various
systematic errors for the $1^{-+}$ state of  1970(90)(300) MeV,
2170(80)(100)(100) MeV and 4390(80)(200) MeV for $n \bar{n}$,  $s
\bar{s}$ and $c \bar{c}$ quarks respectively. For the  $0^{+-}$
spin-exotic they have a noisier signal but evidence that it is heavier.
They also explore mixing matrix elements between spin-exotic hybrid 
states and 4 quark operators.

\section{FULL QCD}

So far I have  discussed  the glueball spectrum, interquark potentials 
and $\alpha_S$ in the quenched approximation. This corresponds 
to treating the sea quarks as of infinite mass (so they don't 
contribute to the vacuum).  To make direct comparison with experiment, 
it is necessary to estimate the corrections from these dynamical quark 
loops in the vacuum. 

The strategy is to use a finite sea-quark mass  but still a value larger
than the empirical light quark mass. The  reason is computational: the
algorithms become very inefficient as the  sea-quark mass is reduced.
The target is to study the effects as  the sea-quark mass is reduced and
then extrapolate to the  physical value.  The present situation,  in
broad terms, is that there is no significant change as the sea-quark
mass  is reduced. This could be because there are no corrections to  the
quenched approximation. Alternatively, the corrections may only turn  on
at a much lower quark mass than has been explored so far.

Let us try to make this argument a little more quantitative. For heavy
sea-quarks of mass $m$, their contribution will be approximately 
proportional to $e^{-2m/E}$ where $E$ is a typical hadronic energy scale
(a few hundred MeV). Thus the quark loop contributions will be
negligible for $m >> E$, which corresponds to the  quenched
approximation. As $m \approx E$, the effects will turn on in a 
non-linear way.

The computational overhead of full QCD on a lattice is  so large because
the quark loops effectively introduce a long range  interaction. The
quark interaction in the action is quadratic  and so can be integrated
out analytically --- see eq(5). This leaves an effective  action  for
the gluonic fields which couples together the fields at all sites.  This
implies that, in a Monte Carlo method, a change in gluon field  at one
site involves the evaluation of the interaction with all other sites. In
practice, one makes small changes at all sites in parallel, but  this
still amounts to inverting a large sparse matrix for each  update. This
is computationally slow.

As the sea-quark mass becomes small, one would expect to need a larger
lattice  size to hold the quarks. For heavy quarks,  the effective range of
the quark loops  in the vacuum will be of order $1/m$. Thus the quenched
approximation  corresponds to $m \to \infty$ and a local interaction. 
For light quarks of a few MeV mass, the range will not be $1/m$, 
because quarks are confined. The lattice studies that have been made 
suggest that spatial sizes of order twice those adequate for the quenched 
approximation are needed for full QCD. This also implies considerable 
computational commitment.

A popular indication of how close a full QCD study is to experiment  is
to ask whether the $\rho$ meson can decay to two pions.  Since the 
decay is P-wave, it needs non-zero momentum. On a lattice  spatial
momentum is quantised in units of $2\pi/L$. Thus  $m_V > 2 ( {m_P}^2 +
4\pi^2/L^2)^{1/2}$ for the decay channel to be allowed energetically. At
present this criterion is rarely satisfied in quenched studies, let 
alone in the more computationally demanding case of full QCD.

The  conclusion of current full QCD lattice calculations is that 
the expected sea-quark effects are not yet fully present.
The main effect observed in full QCD calculations is 
that the lattice parameter $\beta$ which multiplies the gluonic 
interaction term in the action is shifted. Apart from this 
renormalisation of $\beta$, there is little sign of any other 
statistically significant non-perturbative effect.

Consider the changes to be expected for the  inter-quark potential when
the full QCD vacuum is used:
 \begin{itemize}
 \item At small separation $R$, the quark loops will increase the 
size of the effective coupling $\alpha$ compared to the pure gluonic case.
This effect can be estimated in perturbation theory and the change 
at lowest order will be from 1/33 to 1/(33-$2N_f$).
 \item At large separation $R$, the potential energy will saturate at 
a value corresponding to two `heavy-quark mesons'. In other 
words, the flux tube between the static quarks will break by the 
creation of a $q \bar{q}$ pair from the vacuum. 
 \end{itemize}
 Current lattice simulation\site{sesamv} shows some evidence for the
former effect  but no statistically significant signal for the latter.

\section{OUTLOOK}

Lattice QCD is good for asking `what' not `why'. Lattice results for 
masses and matrix elements are obtained from first principles without 
approximations (except in many cases the quenched approximation is still 
needed to keep the computational resource manageable). No model is 
used but no understanding of the underlying physics is obtained. By 
varying the quark masses, boundary conditions etc, it is possible to 
explore a much wider range of circumstances than is available directly 
from experiment. This is a very valuable tool for validating models 
and learning `why'.

Lattice techniques can extract reliable continuum properties from QCD. 
At present, the computational power available combined with the best
algorithms suffices to give accurate results for many quantities  in the
quenched approximation. The future is to establish accurate  values for
more subtle quantities in the quenched approximation (eg. weak matrix
elements of  strange  particles) and to establish the corrections to the
quenched approximation by full QCD calculations.

I hope that soon we  reach the stage where an experimentalist saying `as 
calculated in QCD' is assumed to be speaking of non-perturbative 
lattice calculations rather than perturbative estimates only.

{\referencestyle

}

\end{document}